\documentclass[aps,preprint]{revtex4}%
\usepackage{amsfonts}
\usepackage{amsmath}
\usepackage{amssymb}
\usepackage{graphicx}%
\providecommand{\U}[1]{\protect\rule{.1in}{.1in}}

\begin{document}
\preprint{ }
\title{Quantum Space-Time Symmetries: \\
A Principle of Minimum Group Representation}
\author{Diego J. Cirilo-Lombardo}
\affiliation{M. V. Keldysh Institute of the Russian Academy of Sciences, Federal Research
Center-Institute of Applied Mathematics, Miusskaya sq. 4, 125047 Moscow,
Russian Federation}
\affiliation{CONICET-Universidad de Buenos Aires, Departamento de Fisica, Instituto de
Fisica Interdisciplinaria y Aplicada (INFINA), Buenos Aires, Argentina.}
\author{Norma G. Sanchez}
\affiliation{International School of Astrophysics Daniel Chalonge - Hector de Vega, CNRS,
INSU-Institut National des Sciences de l'Univers, Sorbonne University  75014
Paris, France.\\
Norma.Sanchez@orange.fr \\
https://chalonge-devega.fr/sanchez
}
\date{\today}
\pacs{PACS number}

\begin{abstract}

We show that, as in the case of the principle of minimum action in classical and quantum mechanics, there exists an even more general
principle in the very fundamental structure of {\it quantum
space-time}: This is the principle of {\it minimal group representation} that allows to
consistently and simultaneously obtain a natural description of the spacetime dynamics and the physical states admissible in it. The theoretical construction is based on the physical states that are average values of the generators of the Metaplectic group $Mp(n)$, the double covering of $SL(2C)$ in a
vector representation, with respect to the {\it coherent states} carrying the spin weight. Our main results here are:  (i) There exists a connection between the dynamics given by the Metaplectic group symmetry generators and the physical states (mappings of the generators
through bilinear combinations of the basic states). (ii) The ground states are coherent states of the Perelomov-Klauder type defined by the action of the
Metaplectic group which divide the Hilbert
space into {\it even} and {\it odd} states mutually orthogonal. They carry a spin weight 1/4 and 3/4 respectively from which, two other basic states can be formed. 
(iii) The physical states, mapped bilinearly with the basic 1/4 and 3/4 spin weight states, plus their symmetric and antisymmetric
combinations, have spin contents $s = 0,\; 1/2, \;1,\; 3/2$ \;and \; $2$. (iv) The generators realized with the bosonic variables of the harmonic
oscillator introduce a natural supersymmetry and a superspace whose line element is the geometrical Lagrangian of our model. 
(v) From the line element as operator level, a coherent physical state of spin 2
can be obtained and naturally related to the metric tensor. 
(vi) The metric tensor is {\it naturally discretized} by taking the discrete
series given by the basic states (coherent states) in the $n$ number representation, reaching the classical (continuous) space-time for $n$ $\rightarrow\infty$. 
(vii) There emerges a relation between the
eigenvalue $\alpha$ of our coherent state metric solution and the black hole area (entropy) as $A_{bh}/4l_{p}^{2} = \left\vert \alpha\right\vert$, 
relating the phase space of the metric found $g_{ab}$ and the
black hole area $A_{bh}$ through the Planck length $l_{p}^{2}$ and the
eigenvalue $\left\vert \alpha\right\vert $ of the coherent states. As a consequence of  the lowest level of the quantum discrete  space-time spectrum, eg the ground state associated to $n = 0$ and its characteristic
length, there exists a minimum entropy related to the black hole history. 

\bigskip

\bigskip

Norma.Sanchez@orange.fr \\
https://chalonge-devega.fr/sanchez

\end{abstract}

\volumeyear{year}
\volumenumber{number}
\issuenumber{number}
\eid{identifier}
\date{today}
\

\maketitle
\tableofcontents

\section{Introduction}

 A key concept 
for a full quantum theory of gravity is quantum space-time as well as for quantum theory in its own. The basic motivation of this paper is
to demonstrate that, as in the case of classical and quantum mechanics in
which the minimum action is the regulating and determining principle, there
is an even more general principle that intervenes in the very and
fundamental structure of quantum space-time: this is the interplay between
dynamics and symmetry or alternatively matter/energy and space-time. The
maximum simplicity to achieve this goal is based on the Metaplectic group $%
Mp\left( n\right) $ which is the double covering of the $S_{p}(2C)$ group,
and which for the illustrative case that we intend to establish here we fix
to $Mp\left( 2\right) $.

\bigskip

We characterize quantum space-time as originating from a mapping $P(G,M)$
between the real space-time manifold $M$ and the quantum phase space
manifold of a group $G$. Once one component of the momentum $P$ operator is
identified with the time $T$, the space-time metric of $M$ is found using
the metric $g_{ab}$ on the phase space group manifold.

\bigskip

The group's compactness or noncompactness determines the metric's signature;
nonetheless, noncompact groups are necessary for the majority of physical
situations of interest because of the real space-time signature and its
hyperbolic structure..

\bigskip

The quantum space-time established from the harmonic oscillator's phase
space represents the more obviously fundamental examples of this
development. In the case of the normal (real frequency) oscillator, refer to
Refs \cite{Sanchez2}, \cite{Sanchez2019}, \cite{Sanchez3}, and the mapping $%
(X,\,P)\rightarrow (X,\,iT)$, or alternatively $(X,\,P)\rightarrow (X,\,T)$
in the situation of the imaginary frequency  ( the inverted oscillator) which appear in many physical examples, in particular in cosmology (e.g; in the propagation eqs of classical and quantum perturbations). The quantum
space-time algebra of non commutative operator coordinates is the quantum
oscillator algebra. The line element arises from the Hamiltonian (Casimir
operator) and its discretization yields the quantum space levels. The zero
point energy yields the new quantum region splitting the light cone origin
because the classical generating lines $X = \pm T$ are replaced by the 
curves $X^{2}-T^{2} = [X,T]$ which are the quantum hyperbolae due to the non-zero space and time
commutators, and generate in particular the quantum light cone.   \cite{NSPRD2021}, \cite{NSPRD2023}

\bigskip

The inverted oscillator is associated to the hyperbolic space-time
structure, while the normal oscillator yields euclidean (imaginary time)
signature (quantum gravitational instantons).

In various forms, the inverted oscillator can be found in a multitude of
fascinating physical scenarios, including black holes, particle physics, and
contemporary cosmology (which includes inflation and dark energy today).
Refs \cite{NSPRD2021}, \cite{NSPRD2023}, \cite{dVSiebert}, \cite{dVSanchez} 
\cite{GuthPi}, \cite{Albrecht}, .

\bigskip

An important point in this principle of minimum group representation is the
description of quantum spacetime symmetries: The
"algebro-pseudo-differential correspondence" plays a key role. This
correspondence establishes that a radical operator (e.g. a Hamiltonian) is
equivalent in the context of the metaplectic description, to a
Majorana-Dirac type operator with internal variables in the oscillator
representation . This correspondence is exemplified in the expression Eq.(%
\ref{id}), Section IV in this paper. This algebraic interpretation is
significant because it allows for a connection with pseudodifferential
operators and semigroup (Fourier-Integral) representations, as shown below. 
\cite{pseu}: 
\begin{equation*}
\begin{array}{ccccc}
&  & 
\begin{array}{c}
algebraic \\ 
interpretation%
\end{array}
&  &  \\ 
& \nearrow &  & \nwarrow &  \\ 
{\large \mathit{pseudo\;differential \;operators}} &  & \longleftrightarrow & 
& 
\begin{array}{c}
semigroup \\ 
(Fourier-Integral\text{ }representations)%
\end{array}%
\end{array}%
\end{equation*}

\medskip

In this theoretical and physical context, the resulting solution consists of two
types: the basic state and the observable
physical state,which is bilinear with respect to the basic state (e.g.,the mean
value). The basic state is a coherent state corresponding to the Metaplectic
group, which is the double covering of the SL(2C) group, \cite{pseu}-\cite{epjc}%
.

\medskip

\begin{figure}[ptb]
\begin{center}
\includegraphics[
height=12.263000in,
width=8.690500in,
height=6.4749in,
width=4.5999in
]{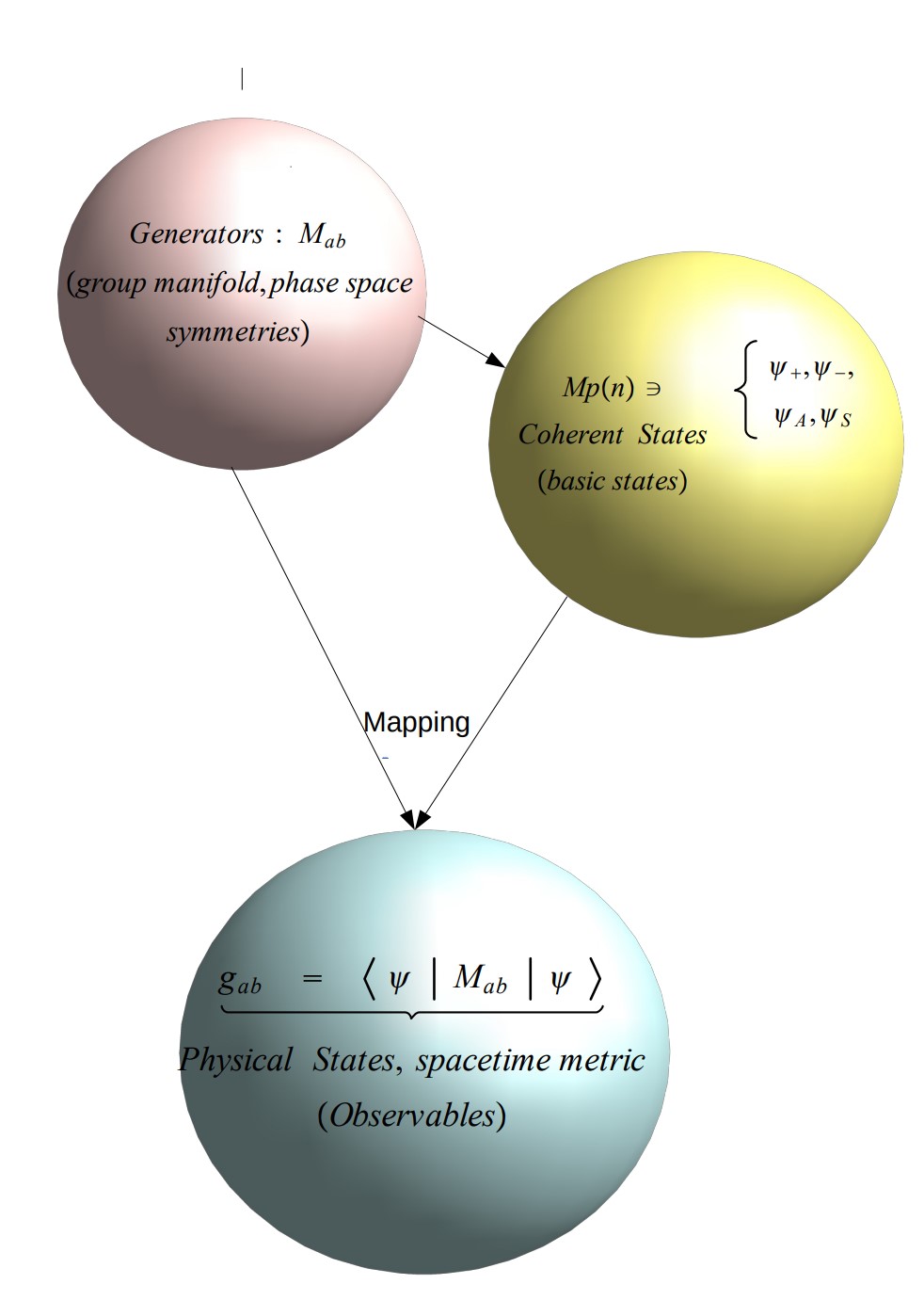}
\end{center}
\end{figure}

We use as our example, Ref \cite{pseu}, a N = 1 superspace with an
invertible and nondegenerate supermetric, where the unconstrained
quantization is precisely carried out using novel techniques based on
coherent states and keeping the Hamiltonian's form. Thus, from the discrete
spectrum of the states themselves, a discrete structure of the spacetime
automatically arises without any prescription of discretization.

\bigskip

Due to the Metaplectic representation (double covering of the $SL(2C)$) of
the coherent state solution representing the emergent spacetime, the
crossover from the quantum microscopic regime to the macroscopical regime
(classical or not) is natural and consistent. This important fact allows us
to conciliate apparently different pictures as that of a macroscopical
quantum gravity regime and that of a dynamical quantum microscopic picture
(the complete process of black hole emission in all its stages being a clear
example).

\bigskip

Despite its simplicity, the framework introduced here have provided
physically and geometrically significant responses concerning an accurate
description of quantum gravity.

\bigskip

It is convenient to think of this kind of coherent states as arising from a
Lie group $G$ operating on a Hilbert space $\mathcal{H}$ through a unitary,
irreducible representation $T$. The set of vectors $\psi \in \mathcal{H}$
such that $\psi =T(g)\,\psi _{0}$ for some $g\in G$. is what we describe as
the coherent state system$\{T,\psi _{0}\}$ for a fixed vector $\psi _{0}$.
We define the states $\left\vert \psi \right\rangle $ corresponding to these
vectors in $\mathcal{H}$ as generalized coherent states.

\section{The Metaplectic group and the principle of minimal representation}

\subsection{Mp(2), SU(1,1) and Sp(2)}

We briefly describe now the relevant symmetry group to achieve the
realization of the Hamiltonian operator of the problem. Specifically, this
group is the metaplectic $Mp\left( 2\right) $ group as the groups $SU(1,1)$ and $%
Sp(2)$ that are topologically covered by it. In function of the $(q,\,p)$
operators, or equivalently the operators $(a,\,a^{+})$ of the standard
harmonic oscillator, the generators of $Mp\left( 2\right) $ are: 
\begin{align}
T_{1}& \,=\,\frac{1}{4}\left( \,q\,p+p\,q\,\right) \,=\,\frac{i}{4}\left(
\,a^{+2}-a^{2}\,\right) ,  \label{cr} \\
T_{2}& \,=\,\frac{1}{4}\left( \,p^{2}-q^{2}\,\right) =-\,\frac{1}{4}\left(
\,a^{+2}+a^{2}\,\right) ,  \notag \\
T_{3}& \,=\,-\frac{1}{4}\left( \,p^{2}+q^{2}\,\right) \,=\,-\frac{1}{4}%
\left( \,a^{+}a+a\,a^{+}\,\right)  \notag
\end{align}%
With the commutation relations,%
\begin{equation*}
\left[ \,T_{3}\,,\,T_{1}\,\right] \;=\;i\,T_{2}\,;\qquad \left[
\,T_{3}\,,\,T_{2}\,\right] \;=\;-\,i\,T_{1}\,;\qquad \left[ \,T_{1},T_{2}\,%
\right] \;=\;-\,i\,T_{3}
\end{equation*}%
The commutation relations can be written as: $\left[ \,T_{3}\;,\;T_{1}\;\pm
\;iT_{2}\,\right] \,=\,\pm \,\left( T_{1}\pm \;i\,T_{2}\right) ;$ $\left[
\,T_{1}+i\,T_{2}\,,\;T_{1}-iT_{2}\,\right] \,=-\,2\,T_{3}$, then it is easy
to see that: $\,T_{1}+i\,T_{2}\,=\,-\,\frac{i}{2}\,a^{2}$ and $%
T_{1}\,-\,i\,T_{2}\,=\,\frac{i}{2}\,a^{+2}$. Therefore, the oscillator
states $\left\vert \,n\,\right\rangle $ of the number operator are
eigenstates of the $T_{3}$ generator 
\begin{equation*}
T_{3}\;\left\vert \;n\;\right\rangle \;=\;-\;\frac{1}{2}\left( n+\;\frac{1}{2%
}\right) \left\vert \;n\;\right\rangle
\end{equation*}

\section{ The Mp(2) vector representation and its coverings}

The commutation relation that specifies the generators $L_{i}$ is the main
feature of the specific representation that was introduced in [2] $:$

\begin{equation}
\left[ \,L_{i},a^{\alpha }\,\right] \,=\,\frac{1}{2}\,a^{\beta }\left(
\sigma _{i}\right) _{\beta }^{\text{ \ }\alpha }  \label{50}
\end{equation}%
The representation above is a non-compact Lie algebra with the following
matrix form:

\begin{equation}
\sigma _{i}=i\left( 
\begin{array}{cc}
0 & 1 \\ 
1 & 0%
\end{array}%
\right) ,\quad \sigma _{j}=\left( 
\begin{array}{cc}
0 & 1 \\ 
-1 & 0%
\end{array}%
\right) ,\quad \sigma _{k}=\left( 
\begin{array}{cc}
1 & 0 \\ 
0 & -1%
\end{array}%
\right) ,  \label{51,52,53}
\end{equation}%
that obey, in a geometrical way: 
\begin{equation}
\sigma _{i}\wedge \sigma _{j}=-i\sigma _{k},\qquad \sigma _{k}\wedge \sigma
_{i}=i\sigma _{j},\qquad \sigma _{j}\wedge \sigma _{k}=\,i\sigma _{i}
\label{56}
\end{equation}%
We want to remark is the following equivalence :

The generators in the representation of Eq.(\ref{50}) fulfil the relation:%
\begin{equation}
L_{i} \,= \,\frac{1}{2}\,a^{\beta}\,\left( \sigma_{i}\right) _{\beta }^{%
\text{ \ }\alpha}a_{\alpha}\,=\,T_{i}  \label{57}
\end{equation}
where T$_{i}$ are the Metaplectic generators namely [6],[7]:%
\begin{align}
T_{1} & \; = \;\frac{i}{4} \;\left( a^{+2}\,-\,a^{2}\right)  \label{58} \\
T_{2} & \; = \;-\;\frac{1}{4}\;\left( a^{+2}\, + \,a^{2}\right)  \label{59}
\\
T_{3} & \; = \; -\;\frac{1}{4}\;\left( aa^{+}\,+\,a^{+}a\right)  \label{60}
\end{align}

\textbf{Proof}: We can write the generators $L_{i}$ in matrix form as%
\begin{equation}
L_{i}\,=\,\overline{u}\;\mathbb{M}_{i}\;v  \label{61}
\end{equation}%
\begin{equation*}
\overline{u}\,\equiv \,\left( 
\begin{array}{cc}
a^{+} & a%
\end{array}%
\right) ,\qquad v\,\equiv \,\left( 
\begin{array}{c}
a \\ 
a^{+}%
\end{array}%
\right)
\end{equation*}%
The representation Eq. (\ref{50}) is faithful, we take into account that $%
\sigma _{k}$ enter as a "metric" in the sense given in Ref \cite{sann}, that
is, it introduces the signature in the quadratic terms in $a$ and $a^{+}$
Eq.(\ref{61})\ explicitly giving rise to the expression Eq.(\ref{57}).
Therefore, we have:

\begin{align}
M_{1} & \;=\;\frac{i}{4}\,\left( 
\begin{array}{cc}
0 & 1 \\ 
-1 & 0%
\end{array}
\right) \;=\;\frac{1}{4}\;\sigma_{k}\;\sigma_{i}\,\quad  \label{64} \\
M_{2} & =\;-\;\frac{1}{4}\;\left( 
\begin{array}{cc}
0 & 1 \\ 
1 & 0%
\end{array}
\right) \;=\;-\;\frac{1}{4}\;\sigma_{k}\;\sigma_{j}\;\quad  \label{65} \\
M_{3} & \;=\;-\;\frac{1}{4}\;\left( 
\begin{array}{cc}
1 & 0 \\ 
0 & 1%
\end{array}
\right) =-\;\frac{1}{4}\;\sigma_{k}^{2}  \label{66}
\end{align}
Consequently, and by inspection, Eq.(\ref{50}) \ coincides with Eq.(\ref{61}%
): Thus, the equivalence Eq.(\ref{57}) \ is proved.

\section{Symmetry and Dynamics Principle: Steps to follow}

A fundamental component of the dynamic description is the square root type
Hamiltonian or Lagrangian, which is, in theory, a non-local and non-linear
operator. This is because the right physical spectrum is generated by the
invariance under reparametrizations both as a Lagrangian and as a
corresponding Hamiltonian. The fundamental principles of our strategy here
are based on certain elements that are explicitly mentioned in the sequel:

\subsubsection{The invariant action}

{(i) }Considering the space-time-matter structure, the geometric Lagrangian
(functional action) of the theory is the elementary distance function, which
is defined as the positive square root of the line element.

At least, the line element's symmetry matches that provided by the
super-Poincar\'e or Cartan-Killing form of Osp(1,2), enabling a bosonic
realization based on the $a$ and $a^+ $ operators of the conventional
harmonic oscillator. This leads to the metric being non-degenerate and
having extra odd (fermionic) coordinates.

\subsubsection{Extended Hamiltonian of the system}

The geometric Hamiltonian, which is the fundamental classical-quantum
operator, is obtained from (i) in the conventional manner.

\bigskip

From the perspective of the physical states, this universal Hamiltonian
(square root Hamiltonian) has an enlarged phase space because it includes a
zero moment {$P_{0}$} characteristic of the entire phase space at its
highest level.

Time "disappears" from the dynamic equations in a proper time system when
the evolution coincides with the time coordinate. This is prevented by
including a zero momentum {$P_{0}$}, which would otherwise lead to the
arbitrary nullification of the Hamiltonian.

\subsubsection{Relativistic wave equation and the algebraic interpretation}

{(iv) The Hamiltonian }$\mathcal{H}_{s}${, rewritten in differential form,
defines a new relativistic wave equation of second order and degree $1/2$
(square root form). This fact can be reinterpreted as a Dirac-Sudarshan type
equation of positive energies and internal variables (e.g. oscillator type
variables) contained as components of the auxiliary or internal vector }$%
\,L_{\alpha }${:}%
\begin{equation}
\mathcal{H}_{s}\,\Psi \;\equiv \;\sqrt{\;\mathcal{F}\;}\;\left\vert \;\Psi
\;\right\rangle \;\;\leftrightarrow \;\,\;\left\{ \left[ \;\mathcal{F}\;%
\right] _{\beta }^{\alpha }\;\,L_{\alpha }\right\} \,\ \Psi ^{\beta }
\label{id}
\end{equation}%
{\ having the basic solution-states of the system, a para-Bose or para-Fermi
interpretation of } $\left\vert \;\Psi \;\right\rangle ${.} This gives rise
to the main justification for an algebraic interpretation of the radical
operator: we have a clean action at operator level and a consistent number
of states of the system (the Lagrange multiplier method eliminates the
square root in a non physical way doubling the spectrum of physical states).

\subsubsection{Basic states of representation and the spectrum of physical
states}

The basic states{\ }$\left\vert \;\Psi_{s}\;\right\rangle ${\ belong to the
group Mp(n) and have a spin weight s = 1/4, 3/4 in the simplest case Mp(2):
They contain \textit{even} and \textit{odd} sectors (s = 1/4, 3/4) in the
number of levels of the Hilbert space respectively and therefore, they span
non-dense irreducible spaces. }

\medskip

In this way, states that are bilinear in
fundamental functions (corresponding to $\left\vert \;\Psi
_{s=1/4,\;3/4}\;\right\rangle ${\ , form the full physical spectrum. In the case of the Metaplectic group $Mp\left( 2\right) $, these fundamental functions
are $f_{1/4}$ and $f_{3/4}$, having a spin weight $s\,=\,1/4$ and $3/4$
respectively.  A physical state characteristic of $Mp\left( 2\right) $
is given by $\Phi _{\mu }\;=\;\left\langle \,s\,\right\vert $ $L_{\mu
}\,\left\vert \,s^{\prime }\,\right\rangle $ with $(s,\,s^{\prime
}\,=\,1/4,\;3/4\,)$, \thinspace\ and \thinspace\ $L_{\mu }$ being the vector
representation of one of the generators of $Mp\left( 2\right) $.}

\medskip

With the Mp(2) interpretation we can also  describe a \textit{complete%
} multiplet spanning \textit{spins} from $(\,0,\,\;1/2,\,\;1,\,\;3/2,\,\;2%
\,).$ This is a consequence of the fact that with the fundamental states and the allowed
vectorial generators, the tower of states is finite and 
 \textit{all} the states involved
are \textit{physical}, as it must be in the physical context.

\bigskip

\section{Statement of the problem}

Geometrically, we take as the starting point the functional action that will
describe the world-line (measure on a superspace) of the superparticle as
follows:

\begin{equation}
S\,=\,\int_{\tau_{1}}^{\tau_{2}}d\tau\,L\left( \,x,\theta,\overline{\theta }%
\,\right) \, = \,-\,m\int_{\tau_{1}}^{\tau_{2}}d\tau\,\sqrt{\overset{\circ }{%
\,\omega_{\mu}}\overset{\circ}{\,\omega^{\mu}}\,+\,{\mathbf{a}}\;\overset{.}{\theta%
}^{\alpha}\overset{.}{\theta}_{\alpha} \,- \,{\mathbf{a}}^{\ast}\overset{.}{%
\overline{\theta}}^{\overset{.}{\alpha}}\overset{.}{\overline {\theta}}_{%
\overset{.}{\alpha}}}  \label{S}
\end{equation}

\medskip

where $\overset{\circ}{\omega_{\mu}}=\overset{.}{x}_{\mu}-i\,(\,\overset{.}{%
\theta}\ \sigma_{\mu}\overline{\theta}-\theta\ \sigma_{\mu}\overset{.}{%
\overline{\theta}}\,)$, and the dot indicates derivative with respect to the
parameter $\tau$, as usual. The above Lagrangian was constructed considering
the line element (e.g. the measure, positive square root of the interval) of
the non-degenerated supermetric

\begin{equation*}
ds^{2}\;=\;\omega^{\mu}\,\omega_{\mu}\;+\;{\mathbf{a}}\,\omega^{\alpha}%
\omega_{\alpha}\;-\;{\mathbf{a}}^{\ast}\,\omega^{\dot{\alpha}}\omega _{\dot{%
\alpha}},
\end{equation*}

where a superspace $(\,1,3\,|1\,)$ is composed by the bosonic term and the
Majorana bispinor , with coordinates $(t,x^{i},\theta ^{\alpha },\bar{\theta}%
^{\dot{\alpha}})$, being the Maurer-Cartan forms of the supersymmetry group
are: $\omega _{\mu }=dx_{\mu }-i\,(\,d\theta \sigma _{\mu }\bar{\theta}%
-\theta \sigma _{\mu }d\bar{\theta}\,),\qquad \omega ^{\alpha }=d\theta
^{\alpha },\qquad \omega ^{\dot{\alpha}}=d\theta ^{\dot{\alpha}}$ , with
evident supertranslational invariance.

\bigskip

As our manifold have extended to include fermionic coordinates, it is
natural to extend also the concept of trajectory for a point particle to the
superspace. Consequently, we take the coordinates $x\left( \tau \right) $, $%
\theta ^{\alpha }\left( \tau \right) $ and $\overline{\theta }^{\overset{.}{%
\alpha }}\left( \tau \right) $ depending on the evolution parameter $\tau .$

\bigskip\ 

The Hamiltonian in square root form, namely $\sqrt{m^{2}-\mathcal{P}_{0}%
\mathcal{P}^{0}-\left( \mathcal{P}_{i}\mathcal{P}^{i}+\frac{1}{a}\Pi
^{\alpha }\Pi _{\alpha }-\frac{1}{a^{\ast }}\Pi ^{\overset{.}{\alpha }}\Pi _{%
\overset{.}{\alpha }}\right) }\left\vert \Psi \right\rangle =0$, is
constructed defining the supermomenta as usual and the Lanczos method for
constrained Hamiltonian systems was used, due the nullification of this
Hamiltonian

\bigskip

Therefore, an algebraic realization of the
pseudo-differential operator (square root) does exist in the case of an underlying Mp$%
\left( n\right) $ group structure:

\begin{equation}
\mathcal{\sqrt{H}}\,\left\vert \Psi\right\rangle \;\equiv\;\sqrt {m^{2}-%
\mathcal{P}_{0}\mathcal{P}^{0}-\left( \mathcal{P}_{i}\mathcal{P}^{i}+\frac{1%
}{a}\Pi^{\alpha}\Pi_{\alpha}-\frac{1}{a^{\ast}}\Pi^{\overset{.}{\alpha}}\Pi_{%
\overset{.}{\alpha}}\right) }\;\left\vert \Psi\right\rangle \,=\,0
\label{661}
\end{equation}%
\begin{equation}
\left\{ \mathcal{\left[ \,H\,\right] }_{\beta}^{\alpha}\,\left( \Psi
L_{\alpha}\,\right) \right\} \Psi^{\beta}\;\equiv\;\left\{ \left[ m^{2}-%
\mathcal{P}_{0}\mathcal{P}^{0}-\left( \mathcal{P}_{i}\mathcal{P}^{i}+\frac{1%
}{a}\Pi^{\alpha}\Pi_{\alpha}-\frac{1}{a^{\ast}}\Pi^{\overset{.}{\alpha}}\Pi_{%
\overset{.}{\alpha}}\right) \right] _{\beta}^{\alpha }\left( \Psi
L_{\alpha}\right) \right\} \Psi^{\beta}\,=\,0  \label{662}
\end{equation}

\bigskip Therefore, both structures can be identified: e.g. $\mathcal{\sqrt {%
H}}\leftarrow\rightarrow\mathcal{\left[ \,H\,\right] }_{\beta}^{\alpha
}\,\left( \Psi L_{\alpha}\,\right) $, being the state $\Psi$ the square root
of a spinor $\Phi$ (on which the "square root" Hamiltonian operates) in such a manner
that it can have the bilinear expression $\Phi=\Psi L_{\alpha}\Psi.$

\bigskip

Equation and Eq. (\ref{661}) in the context of our work has its equivalent
second order Dirac-Like operator in the expression given by Equation and Eq.(%
\ref{662}). This type of operator has been developed by Majorana, Dirac
(e.g. \cite{dirac}, \cite{majo}, and others \cite{gsw} ) containing internal
variables of the harmonic oscillator type, and in our original and
particular case, it gives an algebraic interpretation to the radical
operator, with two fundamental objectives fulfilled: interpreting the action
of the square root operator, and describing the relationship between the
physical (bilinear) states and the fundamental (basic) states, as described
in detail in Section VI here.

\medskip

Equation (\ref{662}) is nothing more and nothing less than the algebraic
interpretation of the radical operator: a Majorana Dirac type operator, that
is to say, a equation with internal variables in the sense of Dirac,
Majorana, and others refs, e.g. \cite{dirac}, \cite{majo}, \cite{gsw} with
different spinorial decomposition structure. The curly brackets in Equation
(16) define the limit of the equivalence with the radical operator
expression given by Eq (\ref{661}).

\bigskip

The key observation here is that the operability of the pseudo-differential
"square root" Hamiltonian can be clearly interpreted if it acts on the
square root of the physical states. 
The square root of a spinor certainly exist in the case of the Metaplectic group,\cite{sann}, \cite{arv}, \cite%
{dirac}, \cite{majo} making our interpretation Eq. (\ref{661}) and Eq. (\ref%
{662}) fully consistent from both the relativistic and group theoretical
viewpoint.

\bigskip

Regarding Equation (\ref{662}) we want to emphasize that our paper refers to
the role of the generator of the metaplectic group both in the dynamics and
in the physically admissible states of the model. We stress that the
variables of the harmonic oscillator are internal from the point of view of
the equations, and the origin of these variables is the faithful and
fundamental representation of the symmetry of the generators of the dynamics
of the spacetime through the physical states, such as the mappings of the
generators in that particular representation, as explained in the paper.

\bigskip

The concept and underlying logic of Equations (\ref{661}) and (\ref{662})
are clear: Quantum symmetries contain - give rise to- the classical
structure. The physical states, as well as the metric (spin 2) are emergent
under the action of the symmetry operator via Equation (\ref{661}).
(Moreover, concrete examples of this concept can be found in Ref. \cite%
{cirilo-sanchez} by these authors). Spin and supersymmetry do not need the
Minkowskian structure. The clear example can be seen for the case of spin 2,
in Sections VIII, IX and X of this paper.

Our Eq. (\ref{662}) here is
fully relativistic and capable of including a complete (super) multiplet
spanning spins from $\; 0,\; 1/2,\;1,\;3/2,\;2$ of physical states.

\bigskip

In the next paragraph, we will describe these states (truly spinorial and
relativistic ones) coming from the algebraic correspondence.

\section{Physical states from Symmetries}

Generators (dynamical symmetries) being into a oscillator-like vector
representation (spinorial) are mapped through their mean values with respect
to the basic states (the Mp(n) coherent states) giving rise to the
observable physical states. That is to say, there is an interrelation
between symmetries and physical states. This gives rise to the first
important consequence that, taking into account the unobservable basic
states, the bilinear states that are observable can only contain spins $%
(0,\; 1/2, \;1, \;3/2, \;2)$.

\medskip

Next, we will provide a brief theoretical justification to the above
construction and then, in the following Section, we describe the emergent
space-time discretization mechanism.

\bigskip

It can be noticed that the family of representations can be increased, e.g. as those of the
Hilbert space operators in the  Weyl representation for a great variety of groups, and asymmetric representations of
various forms. In our case here, the big group involved is the \textit{Metaplectic group} $%
Mp\left( 2\right) $ (the covering group of $SL(2C)$). This important group $%
Mp\left( 2\right) $ is also closely related with the para-Bose coherent states and squeezed states (CS and SS).

Let us consider the concept of generalized coherent states (CS)
based in a Lie group $G$ acting on a Hilbert space H through a unitary,
irreducible representation $T$ in the following. The coherent state system $%
\{T,\psi _{0}\}$ is defined as the set of vectors $\psi \in \mathcal{H}$
such that $\psi =T(g)\,\psi _{0}$ for some  $g\in G$, given a fixed vector  $%
\psi _{0}$. These vectors' equivalent states in H are known as generalized
coherent states (states $\left\vert \psi \right\rangle $).\bigskip 

The following coherent state reproducing Kernel for any operator A (not
necessarily bounded) serves as the foundation for our analysis: 
\begin{equation}
K_{\widehat{A}}\left( \alpha ,\alpha ^{\prime };g\right) \;=\;e^{\,\left[
\;\left\vert \alpha \right\vert ^{2}\,-\,\left\vert \alpha ^{\prime
}\right\vert ^{2}\,\right] }\,\left\langle \,\alpha \,\left\vert
\;A\;\right\vert \,\alpha ^{\prime }\,\right\rangle   \label{31}
\end{equation}%
where $\alpha $ and $\alpha ^{\prime }$ are complex variables that
characterize a respective coherent state, and $g$ is an element of $Mp\left(
\,2\,\right) $. The possible \textit{basic CS states} are classified as:%
\begin{equation*}
\left\vert \,\Psi _{1/4}\,\left( t,\xi ,q\right) \,\right\rangle
\;=\;f\left( \xi \right) \,\left\vert \,\alpha _{+}\left( t\right)
\,\right\rangle 
\end{equation*}%
\begin{equation}
\left\vert \,\Psi _{3/4}\,\left( t,\xi ,q\right) \,\right\rangle
\;=\;f\left( \xi \right) \,\left\vert \,\alpha _{-}\left( t\right)
\,\right\rangle   \label{32a}
\end{equation}%
with the following independent, non-equivalent, \textit{symmetric} and 
\textit{anti-symmetric} combinations%
\begin{equation*}
\left\vert \,\Psi ^{S}\,\right\rangle \;=\;\frac{f\left( \xi \right) }{\sqrt{%
2}}\,\left( \,\left\vert \,\alpha _{+}\,\right\rangle \,+\,\left\vert
\,\alpha _{-}\right\rangle \,\right) \;=\;f\left( \xi \right) \,\left\vert
\,\alpha ^{S}\,\left( t\right) \,\right\rangle 
\end{equation*}%
\begin{equation}
\left\vert \,\Psi ^{A}\,\right\rangle \;=\;\frac{f\left( \xi \right) }{\sqrt{%
2}}\,\left( \,\left\vert \,\alpha _{+}\,\right\rangle \,-\,\left\vert
\,\alpha _{-}\right\rangle \,\right) \;=\;f\left( \xi \right) \,\left\vert
\,\alpha ^{A}\,\left( t\right) \,\right\rangle   \label{32b}
\end{equation}%
The important fact in order to evaluate the kernels Eq. (\ref{31}) is the
action of $a$ and $a^{2}$ over the states previously defined%
\begin{equation*}
a\left\vert \Psi _{1/4}\right\rangle =\alpha \left\vert \Psi
_{3/4}\right\rangle ;\;\;a\left\vert \Psi _{3/4}\right\rangle =\alpha
\left\vert \Psi _{1/4}\right\rangle ;\;\;a\left\vert \Psi ^{S}\right\rangle
=\alpha \left\vert \Psi ^{S}\right\rangle ;\;\;a\left\vert \Psi
^{A}\right\rangle =-\alpha \left\vert \Psi ^{A}\right\rangle 
\end{equation*}%
\begin{equation*}
a^{2}\left\vert \Psi _{1/4}\right\rangle =\alpha ^{2}\left\vert \Psi
_{1/4}\right\rangle ;\;\;a^{2}\left\vert \Psi _{3/4}\right\rangle =\alpha
^{2}\left\vert \Psi _{3/4}\right\rangle ;\;\;a^{2}\left\vert \Psi
^{S}\right\rangle =\alpha ^{2}\left\vert \Psi ^{S}\right\rangle
;\;\;a^{2}\left\vert \Psi ^{A}\right\rangle =\alpha ^{2}\left\vert \Psi
^{A}\right\rangle 
\end{equation*}%

\medskip

and similarly for the states $\overline{\Psi }$ .

\medskip

We have that the \textit{physical states} are particular representations of
the operators $L_{ab}$ and $\mathbb{L}_{ab}$ $\in Mp\left( 2\right) $ in
spinorial form in the sense of quasi-probabilities (tomograms in the $\Psi
_{s}$ plane) or as mean values with respect to the basic coherent states Eqs
(\ref{32a}), (\ref{32b}): $\left\vert \Psi_{\lambda}\right\rangle $,\, $%
\,\lambda\;=\; \left( 1/4,\;1/2,\;3/4,\;1\,\right) $. There are six possible
generalized kernels Eq. (\ref{31}): Two $g\,(\,t,\;s,\pm \;\alpha\,),\;s =
1,\,2$ \thinspace\ in the Heinsenberg Weil (HW) oscillator representation
corresponding to the symmetric and anti-symmetric states respectively: 
\begin{equation}
\left. g_{ab}\,\left( \,t,\,2,\,\alpha\,\right) \,\right\vert
_{HW}\;=\;\left\langle \;\Psi^{S}\left( t\right) \;\right\vert
L_{ab}\;\left\vert \;\Psi^{S}\;\left( \,t\right) \;\right\rangle \;=\;%
\mathcal{F}\;\left( 
\begin{array}{c}
\alpha \\ 
\alpha^{\ast}%
\end{array}
\right) _{\left( 2\right) \,ab}  \label{m}
\end{equation}%
\begin{equation}
\left. g_{ab}\,\left( \,t,\,1,\,-\,\alpha\,\right) \,\right\vert
_{\,HW}\;=\;\left\langle \;\Psi^{A}\left( t\right) \;\right\vert
\,L_{ab}\,\left\vert \;\Psi^{A}\left( t\right) \;\right\rangle \;\,=\,\;%
\mathcal{F}\;\left( 
\begin{array}{c}
-\alpha \\ 
-\alpha^{\ast}%
\end{array}
\right) _{\left( 1\right) \,ab}  \label{n}
\end{equation}
where : 
\begin{equation*}
\mathcal{F}\; = \;e\,^{[\,-\,\left( \, \frac{m}{\sqrt{\,2}\;\left\vert \, 
\mathbf{a}\,\right\vert }\,\right) ^{2}\,\left[ \;\left( \,\alpha
\,+\,\alpha^{\ast}\,\right) \,-\,B\;\right] ^{2}\,+\,D\;]}\,e\,^{[\;\xi
\,\varrho\;\left( \,\alpha\,+\,\alpha^{\ast}\,\right) \;]}\;\left\vert \,f
\left( \, \xi\,\right) \,\right\vert ^{2}
\end{equation*}

and four $g_{ab}\,(\,t,\,s,\,\alpha^{2}\,)$ , $s=(\,1,\;2,\;1/2,\;3/2\,)$
for $SU(1,1)$, with the symmetric $\Psi^{S}$, anti-symmetric $\Psi^{A}$, and 
$\Psi_{1/4}$, $\Psi_{3/4}$ states: 
\begin{equation}
g_{ab}\left( \,t,\,2, \,\alpha^{2}\,\right) _{SU(1,1)} = \;\left\langle
\Psi^{S}\left( t\right) \right\vert \,\mathbb{L}_{ab}\,\left\vert \Psi
^{S}\left( t\right) \right\rangle \;=\;\mathcal{F}\;\left( 
\begin{array}{c}
\alpha^{2} \\ 
\alpha^{\ast2}%
\end{array}
\right) _{\left( 2\right) \,ab}  \label{a}
\end{equation}%
\begin{equation}
g_{ab}\left( \,t,\,1, \,\alpha^{2}\,\right) _{SU(1,1)} = \;\left\langle
\Psi^{A}\left( t\right) \, \right\vert \mathbb{L}_{ab}\left\vert \Psi
^{A}\left( t\right) \right\rangle \;=\;\mathcal{F}\;\left( 
\begin{array}{c}
\alpha^{2} \\ 
\alpha^{\ast2}%
\end{array}
\right) _{\left( 1\right) \,ab}  \label{b}
\end{equation}%
\begin{equation}
g_{ab}\left( \,t,\,3/2,\,\alpha^{2}\,\right) _{SU(1,1)}\;=\;\left\langle
\Psi_{3/4}\left( t\right) \right\vert \mathbb{L}_{ab}\left\vert \Psi
_{3/4}\left( t\right) \right\rangle \;=\;\mathcal{F}\;\left( 
\begin{array}{c}
\alpha^{2} \\ 
\alpha^{\ast2}%
\end{array}
\right) _{(3/2)ab}  \label{c}
\end{equation}%
\begin{equation}
g_{ab}\left( \,t,\,1/2,\,\alpha^{2}\,\right) _{SU(1,1)}\;=\;\left\langle
\Psi_{1/4}\left( t\right) \right\vert \mathbb{L}_{ab}\left\vert \Psi
_{1/4}\left( t\right) \right\rangle \; = \;\mathcal{F}\;\left( 
\begin{array}{c}
\alpha^{2} \\ 
\alpha^{\ast2}%
\end{array}
\right) _{(1/2)\,ab}  \label{33}
\end{equation}
where $B$ and $D$ are given by: 
\begin{equation}
B\;=\;\left( \,\frac{\left\vert \,\mathbf{a}\right\vert }{m}\,\right)
^{2}\,c_{1}, \quad\;,\;\quad D\;=\;\left( \frac{\left\vert \,\mathbf{a}%
\,\right\vert \,c_{1}}{\sqrt{2}\,m}\,\right) ^{2}\,+\,c_{2}  \label{l}
\end{equation}

$c_{1}$ and $c_{2}$ being constants characterizing the solution or its
initial conditions.

\bigskip

The dynamical structure of (quantum) spacetime clearly encodes the metric
through the coherent basic states, solutions of the Equations (\ref{661})
and (\ref{662}). Therefore, the spacetime structure defined in this paper
through the metrics in Equations (\ref{m})-(\ref{33}) fully and rigorously
respect all the properties required in the fundamental quantum regime, as
well as in the classical domain.

\bigskip

Eqs. (\ref{33}) are expressed in the so called Sudarshan's
diagonal-representation that lead, as an important consequence, the \textit{%
physical states} with spin content $\lambda \;=\;(\,1/2,\;1,\;3/2,\;2\,)$.
Precisely, the generalized coherent states here generate a map that relates
the metric, solution of the wave equation $g_{ab}$ to the specific subspace
of the full Hilbert space where these coherent states\ live. Moreover, there
exists for operators $\in Mp\left( 2\right) $an asymmetric - kernel leading
for our case the following $\lambda=1$ state :%
\begin{equation*}
\left. g_{ab}\left( \,t,\,1,\,\alpha\right) \,\right\vert
_{HW}\;=\;\left\langle \Psi_{3/4}\left( t\right) \right\vert
L_{ab}\left\vert \Psi_{1/4}\left( t\right) \right\rangle =\left\langle
\Psi_{1/4}\left( t\right) \right\vert L_{ab}\left\vert \Psi_{3/4}\left(
t\right) \right\rangle \;=\;\mathcal{F}\;\left( 
\begin{array}{c}
\alpha \\ 
\alpha^{\ast}%
\end{array}
\right) _{\left( 1\right) \,ab}
\end{equation*}
This is so because the non-diagonal projector involved in the reconstruction
formula of $L_{ab}$ is formed with the $\Psi_{1/4}$ and $\Psi_{3/4}$ states
which span completely the \textit{full} Hilbert space.

\bigskip

\textit{Observation 1}: Due to the non observability of isolated basic
states, the spin zero physical states appear as bounded states $(\,g\,%
\overline{g}\,)$, where \thinspace\ $g_{ab}\,\left( \,t,s,w\,\right) $%
\thinspace\ and \thinspace\ $\overline{g}_{ab}\,\left( \,t,s,w\,\right) $
are given by the bilinear expressions Eqs (\ref{33}).

\bigskip

\textit{Observation 2}: Each kernel represents a global \textit{physical}
state composed by fundamental states that separately are \textit{basic} and 
\textit{unobservable}.

\bigskip

Notice that the spectrum of the physical states are labeled not only by
their spin content $\lambda$, but also by the "eigenspinors" $\left( 
\begin{array}{c}
\alpha \\ 
\alpha^{\ast}%
\end{array}
\right)_{(\lambda)\,ab}$ and \thinspace\ $\left( 
\begin{array}{c}
\alpha^{2} \\ 
\alpha^{\ast2}%
\end{array}
\right) _{(\lambda)\,ab}$ corresponding to the vector representations of $%
L_{ab}$ and $\mathbb{L}_{ab}$ respectively, (maps over a region of $\mathcal{%
H}$).

\section{Supermetric and emergent spacetime}

The Lagrangian density from the action Eq.(\ref{S}) represents a free
particle in a superspace with coordinates $z_{A}\equiv\left(
x_{\mu},\theta_{\alpha },\overline{\theta}_{\overset{\cdot}{\alpha}}\right) $%
. In these coordinates, the line element of the superspace reads, 
\begin{equation}
ds^{2}\longrightarrow\,\dot{z}^{A}\dot{z}_{A}\,=\,\dot{x}^{\mu}\dot{x}%
_{\mu}-2 \,i\,\dot{x}^{\mu}\,( \,\dot{\theta}\,\sigma_{\mu}\,\bar{\theta}%
-\theta\,\sigma_{\mu}\dot{\bar{\theta}}\,)+\left( {\mathbf{a \,-\,}}\bar{%
\theta}^{\dot{\alpha}}\bar{\theta}_{\dot{\alpha}}\right) \dot{\theta }%
^{\alpha}\dot{\theta}_{\alpha}-\left( {\mathbf{a}}^{\ast}+ \,\theta^{\alpha
}\theta_{\alpha} \,\right) \dot{\bar{\theta}}^{\dot{\alpha}}\dot{\bar{\theta 
}}_{\dot{\alpha}}  \notag
\end{equation}

It is important to notice that following the steps detailed in Section IV,
the quantization is exactly performed providing the correct physical and
mathematical interpretation to the square root Hamiltonian, and the correct
spectrum of physical states.

\bigskip

Without lose of generality, and for simplicity, we take the solution Eq. (%
\ref{m}) to represent the metric and with three compactified dimensions ( $s
= 2$ spin fixed), we have : 
\begin{equation}
g_{AB}\,(t)\,=\,e^{\,A(t)\,+\,\xi\,\varrho\,(t)}\,g_{AB}\,(0),  \label{CSsol}
\end{equation}
where the initial values of the metric components are given by 
\begin{equation}
g_{ab}\,(0)\,=\,\langle\psi(0)|\left( 
\begin{array}{c}
a \\ 
a^{\dagger}%
\end{array}
\right) _{ab}|\psi(0)\rangle,  \label{g0}
\end{equation}
or, explicitly, 
\begin{align}
g_{\mu\nu}(0) & =\eta_{\mu\nu}\,,\qquad g_{\mu\alpha}(0)= -\, i\,\sigma
_{\mu\alpha\dot{\alpha}}\bar{\theta}^{\dot{\alpha}}\,,\qquad g_{\mu\dot {%
\alpha}}(0) = - \,i\,\theta^{\alpha}\sigma_{\mu\alpha\dot{\alpha}}\,,
\label{gg} \\
g_{\alpha\beta}(0) & = (a-\bar{\theta}^{\dot{\alpha}}\bar{\theta}_{\dot{%
\alpha}})\,\epsilon_{\alpha\beta}\,,\qquad g_{\dot{\alpha}\dot{\beta}}(0) =
-\,(a^{\ast} +\theta^{\alpha}\theta_{\alpha})\,\epsilon_{\dot{\alpha }\dot{%
\beta}}\,.  \label{gdiego}
\end{align}

The bosonic and spinorial parts of the exponent in the superfield solution
Eq. (\ref{CSsol}) are, respectively, 
\begin{equation}
\begin{array}{rcl}
A(t) & = & -\left( \frac{m}{|{\mathbf{a}}|}\right) ^{2}t^{2} \,+ \, c_{1}\,t
\, + \,c_{2}, \\ 
\xi\,\varrho\left( t\right) & = & \xi\left( \,\phi_{\alpha}\,(t) \,+ \,\bar{%
\chi}_{\dot{\alpha}}\,(t)\,\right) \\ 
& = & \theta^{\alpha}\left( \overset{\circ}{\phi}_{\alpha}\cos\,\left( \,
\omega\, t/\,2 \,\right) \, + \,\frac{2}{\omega} \,Z_{\alpha}\,\right) -%
\overline{\theta}^{\overset{\cdot}{\alpha}}\left( - \,\overset{\circ }{%
\overline{\phi}}_{\overset{\cdot}{\alpha}}\sin\,\left( \,\omega t/2\,\right)
-\frac{2}{\omega}\overline{Z}_{\overset{.}{\alpha}}\right) \\ 
& = & \theta^{\alpha}\overset{\circ}{\phi}_{\alpha}\cos\left( \,\omega t/2
\,\right) \, + \,\bar{\theta}^{\overset{\cdot}{\alpha}}\,\overset{\circ }{%
\overline{\phi}}_{\overset{\cdot}{\alpha}}\,\sin\left( \,\omega
t/\,2\,\right) \, + \,4\,|\,\mathbf{a}\,| \;Re(\,\theta Z\,),%
\end{array}
\label{expo}
\end{equation}
where $\overset{\circ}{\phi}_{\alpha}, \, Z_{\alpha},\,\overline{Z}_{\overset%
{.}{\beta}}$ are constant spinors, $\omega\,= \,1/|\,\mathbf{a}\,|$ and the
constant $c_{1}\in\mathbb{C}$, due to the obvious physical reasons and the
chiral restoration limit of the superfield solution. We see in the next
Section the associated emerging discrete space-time structure.

\section{Superspace and discrete spacetime structure}

Let us see in this Section how the discrete spacetime structure emerges naturally from
the model under consideration. Expanding on a basis of eigenstates of the number operator: 
\begin{equation}
\sum_{m} \;|m\rangle\;\langle m|\;= \;1,  \label{nb}
\end{equation}
we have 
\begin{equation}
g_{ab} \,(0) \,= \,\sum_{n,m}\,\langle\,\psi(0)\,|m\rangle\;\langle
\,m\,|\;L_{ab}\;|\,n \;\rangle\;\langle\,n\,|\,\psi(0)\,\rangle  \label{dg}
\end{equation}
Then, 
\begin{equation*}
g_{ab}(t) \,=\,\underset{f\left( t\right) }{\underbrace{e^{A(t)\,+\,\xi
\,\rho(t)}}} \,\sum_{n,m}\,\langle\psi(0)\,|\,m\rangle\;\langle n\,|\,\psi
(0)\,\rangle\;\langle m|\,L_{ab}\,|\,n\rangle\,
\end{equation*}%
\begin{equation}
\langle m \,|\;L_{ab}\;|\,n\rangle= \langle m|\left( 
\begin{array}{c}
a \\ 
a^{\dagger}%
\end{array}
\right) _{ab}|n\rangle=\left( 
\begin{array}{c}
\langle m|n-1\rangle\sqrt{n} \\ 
\langle m|n+1\rangle\sqrt{n+1}%
\end{array}
\right) _{ab}=\left( 
\begin{array}{c}
\delta_{m,n-1}\sqrt{m} \\ 
\delta_{m,n+1}\sqrt{m+1}%
\end{array}
\right) _{ab}  \label{7}
\end{equation}
It follows%
\begin{equation*}
g_{ab}(0)=\sum_{n,m}\,\langle\,\psi(0)\,|m\rangle\left( 
\begin{array}{c}
\delta_{m,n-1}\sqrt{m} \\ 
\delta_{m,n+1}\sqrt{m+1}%
\end{array}
\right) _{ab}\langle\,n\,|\,\psi(0)\,\rangle
\end{equation*}%
\begin{equation*}
g_{ab}(0)=\sum_{n}\,\sqrt{n}\;\langle\,\psi(0)\,| \,n-1\,\rangle\;\langle\,
n|\,\psi(0)\rangle\,\left( 
\begin{array}{c}
1 \\ 
0%
\end{array}
\right) _{ab}+\;\;\sum_{m}\sqrt{n+1}\;\langle\,\psi(0)\,|n+1\rangle\;\langle
n|\,\psi(0)\,\rangle\left( 
\begin{array}{c}
0 \\ 
1%
\end{array}
\right) _{ab}
\end{equation*}

\medskip

According to the equation above, the splitting of  $\psi $ into the
fundamental states of the metapletic representation is the only explanation
that makes sense.%
\begin{equation}
|\;\psi (0)\;\rangle \;=\;A\;|\alpha _{+}\rangle +B\;|\alpha _{-}\rangle 
\label{phy}
\end{equation}%
Consequently, at the macroscopic level, the arbitrary constants $A$ and $B$
govern the spectrum's classical behavior. Without losing generality, we
assume for the purposes of this discussion that $A=B$ such that $|\psi
(0)\rangle =|\alpha _{+}\rangle +|\alpha _{-}\rangle .$ However, we will
come back to this crucial point later.

This is the outcome of the $SO(2,1)$ group's breakdown into two irreducible
representations of the metaplectic group $Mp(2)$, spanning even and odd n,
respectively.

\bigskip

Let us highlight the important  property of the \ state $|\,\psi (0)\,\rangle =|\,\alpha
_{+}\,\rangle \;+\;|\alpha _{-}\rangle $ which (if $A = B$) is invariant 
to the action of the operators $a$ and $a^{\dagger }$. This is a consequence of the fact that in  the metaplectic representation the general behaviour of these states are: 
$a\,|\alpha _{+}\rangle =a^{\dag }\,|\alpha _{+}\rangle =|\alpha _{-}\rangle 
$ and $a\,|\alpha _{-}\rangle =a^{\dag }|\alpha _{-}\rangle =|\alpha
_{+}\rangle $.

\subsection{Statistical distributions and classical limit}

From the Poissonian distribution for the coherent states we can see: 
\begin{equation*}
P_{\alpha }(n)=|\,\langle \,n\,|\,\alpha \,\rangle \,|^{2}=\frac{\alpha
^{n}e^{-\alpha }}{n!}
\end{equation*}%
fulfilling%
\begin{equation*}
\underset{n=0}{\overset{\infty }{\sum }}P_{\alpha }(n)=1,\ \ \ \ \underset{%
n=0}{\overset{\infty }{\sum }\,n}\,P_{\alpha }(n)\,=\,\alpha \ 
\end{equation*}

\medskip

It is different from the individual distributions defined from each one of
the two irreducible representations of the metaplectic group Mp(2) (which span
even and odd $n$ respectively):

\begin{equation}
\left. 
\begin{array}{c}
\underset{n=0}{\overset{\infty}{\sum}}P_{\alpha_{+}}(2n)\;=\;
e^{-\alpha}\cosh(\alpha) \\ 
\underset{n=0}{\overset{\infty}{\sum}}P_{\alpha_{-}}(2n+1) \;= \;
e^{-\alpha}\sinh(\alpha)%
\end{array}
\right\} \;\,\rightarrow\; \; \underset{n=0}{\overset{\infty}{\sum}}\left(
P_{\alpha_{+}}(n)\;+\;P_{\alpha_{-}}(n)\right) \; = \; 1
\end{equation}

\medskip

Notice that in despite of the different form between the above
equations, the limit $\ n\rightarrow \infty $ is the same for both: the sum of the
two distributions arising from the Mp$\left( 2\right) $ irreducible
representations (IR), and for the $SO(2,1)$ representation as it must be.

\medskip

Taking this into account, the explicit form of $|\,\alpha _{+}\,\rangle
,\,|\alpha _{-}\rangle $ are given by : 

\begin{align}
|\,\alpha _{+}\,\rangle & \;\equiv \;\left\vert \,\Psi _{1/4}\left( 0,\xi
,q\right) \,\right\rangle \;=\;\overset{+\infty }{\underset{k=0}{\sum }}%
\;f_{2k}\left( 0,\xi \right) \left\vert 2k\right\rangle \;=\;\overset{%
+\infty }{\underset{k=0}{\sum }}\;f_{2k}\left( 0,\xi \,\right) \frac{\left(
a^{\dag }\right) ^{2k}}{\sqrt{\left( 2k\right) !}}\;\left\vert
0\right\rangle   \label{110} \\
|\alpha _{-}\rangle & \equiv \left\vert \,\Psi _{3/4}\,\left( \,0,\xi
,q\,\right) \,\right\rangle \;=\;\overset{+\infty }{\underset{k=0}{\sum }}%
\;f_{2k+1}\left( 0,\xi \right) \;\left\vert 2k+1\right\rangle \;=\;\overset{%
+\infty }{\underset{k=0}{\sum }}\;f_{2k+1}\left( 0,\xi \right) \frac{\left(
a^{\dagger }\right) ^{2k+1}}{\sqrt{\left( 2k+1\right) !}}\left\vert
0\right\rangle   \notag
\end{align}

\medskip

where  all the possible odd $n$ dependence is stored in the parameter $\xi$.

\medskip

Consequently, $|\alpha _{+}\rangle $ connects only with \textit{even}
vectors of the basis number and $|\alpha _{-}\rangle $ with the \textit{odd}
vectors in the basis number. Therefore, using the decomposition Eq. (\ref%
{phy}) and decomposing the base number $\left\vert n\right\rangle $ into 
\textit{even} and \textit{odd}, we obtain the following explicit result for the space-time metric : 

\medskip

{\small 
\begin{gather}
g_{ab}(t)\;=\;\frac{f\left( t\right) }{2}\sum_{m}\left\{ \left[ \;P_{\alpha
_{+}}(2m)\cdot 2m+P_{\alpha -}(2m+1)\cdot \left( 2m+1\right) \;\right]
\left( 
\begin{array}{c}
1 \\ 
0%
\end{array}%
\right) _{ab}+\right.   \label{mp} \\
+\left. \left[ P_{\alpha _{+}^{\ast }}(2m)\cdot 2m+P_{\alpha _{-}^{\ast
}}(2m+1)\cdot (2m+1)\right] \left( 
\begin{array}{c}
0 \\ 
1%
\end{array}%
\right) _{ab}\right\}   \notag
\end{gather}%
} 
\medskip

The expression above is an important pillar of our findings here: In this equation 
\textit{the discrete structure of the spacetime} is shown explicitely as the fundamental basic feature 
of a consistent quantum field theory of gravity.

\medskip

On the other hand, in the limiting case $n\rightarrow\infty$ our solution for metric
 goes to the continuum one as, it must be:%

\begin{equation*}
\underset{n=0}{\overset{\infty}{\sum}}\left[ P_{\alpha_{+}}(2m)\cdot
2m+P_{\alpha-}(2m+1)\cdot(2m+1)\right] =\alpha \;e^{-|\alpha|}\left(
\cosh(\alpha)+\sinh(\alpha)\right) =\alpha
\end{equation*}

Similarly, for the lower part (spinor down) of the above equation, we obtain:%
\begin{equation*}
\underset{n=0}{\overset{\infty}{\sum}}\left[ P_{\alpha_{+}}(2m)\cdot
2m+P_{\alpha_{-}}(2m+1)\cdot(2m+1)\right] =\alpha^{\ast}
\end{equation*}

Therefore, when the number of discrete levels increases, our metric
solution goes to the general relativistic continuum "manifold" behaviour :

\begin{equation}
g_{ab}(t)_{n\rightarrow\infty} \;\rightarrow\;\frac{f\left( t\right) }{2}%
\left\{ \alpha\left( 
\begin{array}{c}
1 \\ 
0%
\end{array}
\right) _{ab} +\; \alpha^{\ast}\left( 
\begin{array}{c}
0 \\ 
1%
\end{array}
\right) _{ab}\right\} = f\left( t\right) \langle\,\psi(0)\,\left( 
\begin{array}{c}
a \\ 
a^{\dagger}%
\end{array}
\right) _{ab}|\,\psi(0)\,\rangle  \label{12}
\end{equation}
as expected.

\section{The Lowest n = 0 Level and its Length}

Is not difficult to see that for the number $n = 0$ the metric solution
takes the value

{\small 
\begin{align}
g_{ab}(t) & \; = \;\frac{f\left( t\right) }{2}\left[ \,P_{\alpha
}\,(1)\left( 
\begin{array}{c}
1 \\ 
0%
\end{array}
\right) _{ab} + \;P_{\alpha_{-}^{\ast}}(1)\left( 
\begin{array}{c}
0 \\ 
1%
\end{array}
\right) _{ab}\,\right]  \label{13} \\
& \; = \;\frac{f\left( t\right) }{2} \;e^{-\left\vert \,\alpha\,\right\vert }%
\left[ \, \alpha\left( 
\begin{array}{c}
1 \\ 
0%
\end{array}
\right) _{ab} + \;\alpha^{\ast}\left( 
\begin{array}{c}
0 \\ 
1%
\end{array}
\right) _{ab}\,\right]  \notag
\end{align}
}

\medskip

This evidently defines an associated characteristic length for the
eigenvalues $\alpha $, $\alpha ^{\ast }$ because the metric axioms in a
Riemannian manifold. In principle, fundamental symmetries as the Lorentz
symmetry can be preserved at this level of discretization due to the
existence of discrete Poincare subgroups of this supermetric.

\section{Implications for the Black hole entropy: A superspace solution}

The black hole entropy,   $S=k_{B}\,A_{bh}\,/4\,l_{P}^{2}$ where $A$ is the
horizon area and $l_{P}\equiv \sqrt{\hslash G/c^{3}}$ is the Planck length
as is well known, was first found by Bekenstein and Hawking \cite{bh} using
thermodynamic arguments of preservation of the first and second laws of
thermodynamics.

\bigskip

Also found by Bekenstein was an information theory proof in which black hole
entropy is treated as the measure of the "inaccessible"
information for an external observer on an actual internal configuration of
the black hole in a given state. Such state is described by the values
of mass, charge, and angular momentum.

\bigskip

From the statistical mechanics viewpoint, the entropy is the mean logarithm
of the density matrix. About this issue, Bekenstein proposed a model of
quantization of the horizon area with the title "Demystifying black hole's
entropy proportionality to area", Ref \cite{bek-atoms}:

\medskip

Following the same reasoning, the horizon is formed by patches or cells of equal area 
$\delta l_{P}^{2}$. Consequently,  the horizon can be considered as endowed with many fundamental degrees
of freedom: one degree per each patch. Therefore, the horizon does appear composed by fundamental patches all having an equal 
 number $\chi $ of quantum states.

\bigskip

Then, the horizon have a  total number of quantum states given by $\Omega _{H}=\chi
^{A_{bh}\,/\,\delta \,l_{P}^{2}}$ and the Boltzmann statistical entropy
due to the horizon is $S\;=\;k_{B}\,\ln \Omega
_{H}\;=\;k_{B}\;\left( \,A_{bh}\,/\,\delta \,l_{P}^{2}\,\right) \ln \chi $:

\medskip

The choice $\delta= 4\,\ln\,\chi$ yields   the expected
thermodynamical black hole Bekenstein's formula.

\medskip

Introducing $\delta$ into the
original black hole entropy formula one obtains the Poisson expression for
the total number of states: 
\begin{equation}
\Omega_{H} \;= \; e^{A_{bh}\;/\;4\;l_{P}^{2}}  \label{14}
\end{equation}
This expression is explicitely \textit{the bridge} with the structure of the
emergent coherent state metric of our approach here. We consider the similar
Poissonian expression for the number of states from $g_{ab}$, namely $%
e^{\left\vert \alpha\right\vert }$, therefore the relation between the coherent state
eigenvalue $\alpha$ corresponding to our coherent state metric solution and
the above equation does appear:%
\begin{equation}
A_{bh}\,/\,4\,l_{P}^{2}\;= \;\left\vert \,\alpha\,\right\vert  \label{15}
\end{equation}
This expression links the black hole area $A_{bh}$ and the phase space of the coherent state solution
metric $g_{ab}$ through \ the Planck length 
$l_{P}^{2}$ and the eigenvalue $\left\vert \alpha\right\vert $
characterizing the coherent states.

\section{Implications for Hawking Radiation}

\begin{itemize}
\item {As is known, the area of the black hole is related to its mass, consequently the black hole mass in our approach here is quantized as well. The emitted radiation from the black hole does appear  because of the quantum jump from one quantized
value of the mass (energy) to a lower quantized value. The decreasing of the black hole mass occurs because of this process.}

\item {Therefore, (and because radiation is emitted at quantized
frequencies corresponding to the differences between energy levels), quantum
gravity implies a discretized emission spectrum for the black hole radiation.
}

\item {The spectral lines can be very dense in macroscopic regimes leading
physically no contradiction with Hawking's prediction of a continuous
thermal spectrum in the semiclassical regime.}

\item {From the point of view of our approach here:}

\item {\ If we now suppose simply that the constants $A$, $B$ in the state
solution eg. Eq. (27), Eq. (35), are different, $A\neq B$ we have : 
\begin{equation*}
|\psi(0)\rangle= A\,|\,\alpha_{+}\,\rangle\,+\,B \,|\,\alpha_{-}\,\rangle,
\end{equation*}
}

\item {Therefore, the thermal (Hawking) spectrum at the
macroscopic or semiclassical level does not appear.}

\item {This fact is clearly explained  because un exact balance between the
superposition of the two irreducible representations of the Metaplectic
group is needed.}

This will lead as a result, non classical states of radiation in the sense
of \cite{21} as can be easily seen putting, for example, one of the constants, $B$ $%
\left( \text{or }A\right)$ equal to zero:

\begin{equation}
g_{ab}(t)\;=\;A\; \; \frac{f\left( t\right) }{2}\;\sum_{m}\left[
P_{\alpha_{+}}(2m)\cdot\left( \,2m\,\right) \,+\,P_{\alpha-}(\,2m+1\,)\cdot
(\,2m+1\,)\,\right] \left( 
\begin{array}{c}
1 \\ 
0%
\end{array}
\right)_{ab}  \label{16}
\end{equation}

\item {Notice that only the up spinor part survives in this case, and the classical thermal limit is not attained. This is so, even in the continuous limit for this case, in which the
number of levels increases accordingly to }

\begin{equation}
g_{ab}(t)\;_{n\,\rightarrow \,\infty}\;\rightarrow\;\frac{f\left( t\right) }{%
2}\;A\;\alpha\; \left( \, 
\begin{array}{c}
1 \\ 
0%
\end{array}
\right)_{ab}\; = \;A\;f\left(t\right) \langle \;\psi(0) \;|\left( 
\begin{array}{c}
a \\ 
0%
\end{array}
\right)_{ab}|\;\psi(0)\;\rangle  \label{17}
\end{equation}

\item {Consequently,  in such a case where $A = 0$, (or $B = 0$), the spectrum will takes
only \textit{even (or odd)} levels becoming evidently non thermal.}
\end{itemize}

Therefore,

\begin{itemize}
\item {In the case $A = B$ the thermal Hawking spectrum is attained at the continuum
classical gravity level eg, the Poissonian behaviour of the distribution is
complete.}

\item {In the cases, with $A\neq B$, the spectrum is non classical,
and quantum properties of gravity are manifest at the macroscopic level.}
\end{itemize}

\section{Concluding remarks}

Here we have demonstrated that there is a principle of minimal group representation
that allows us to consistently and simultaneously obtain a natural
description of the dynamics of spacetime and the physical states admissible
in it.

\bigskip

The theoretical construction is based on the fact that the physical states
are, roughly speaking, average values of the generators of the metaplectic
group Mp(n) in a vector representation, with respect to the coherent states
that are not observable (carrying the weight of spin). Schematically, we
have the following picture where $M_{ab} = \mathbb{L}_{ab}(L_{ab})$:

\begin{equation*}
\begin{array}{ccccc}
&  & \underset{%
\begin{array}{c}
Physical\text{ }\;States,\text{ }spacetime\text{ }metric \\ 
(Observables)%
\end{array}
}{\underbrace{g_{ab}\;\;=\;\;\left\langle \;\psi\;\right\vert
\;M_{ab}\;\left\vert \;\psi\;\right\rangle }} &  &  \\ 
& \nearrow &  & \nwarrow &  \\ 
\underset{%
\begin{array}{c}
Coherent\;\,States \\ 
(basic\text{ }states)%
\end{array}
}{Mp(n)\ni}\left\{ 
\begin{array}{c}
\psi_{+},\psi_{-}, \\ 
\psi_{A},\psi_{S}%
\end{array}
\right. &  & \longleftrightarrow &  & \underset{}{%
\begin{array}{c}
Generators:\text{ }M_{ab} \\ 
(group\text{ }manifold, \\ 
phase\text{ }space\text{ } \\ 
symmetries)%
\end{array}
}%
\end{array}%
\end{equation*}

\bigskip

\bigskip

In summary:

\bigskip

\textbf{(1)} We demonstrate that there is a connection between dynamics,
given by the generators of the symmetries, and the physically admissible
states.

\bigskip

\textbf{(2)} The physically admissible states are mappings of the generators
of the relevant symmetry groups covered by the metaplectic group, in the
simplest case according to the chain: $Mp(2)\supset SL(2R)\supset SO(1,2))$
through a bilinear combination of basic states.

\bigskip

\textbf{(3)} The ground states are coherent states defined by the action of
metaplectic group (Perelomov-Klauder type), these states divide the Hilbert
space into \textit{even} and \textit{odd} states, and are mutually
orthogonal. They carry a weight of spin 1/4 and 3/4 respectively.

\bigskip

\textbf{(4)} From the basic states combined symmetrically and
antisymmetrically, two other basic states can be formed. These new states
manifest a change of sign under the action of the creation operator $a^+$.

\bigskip

\textbf{(5)} The physically admissible states, mapped bilinearly with the
basic states with spin weight 1/4 and 3/4, plus their symmetric and
antisymmetric combinations, have spin contents s\; = \;0,\; 1/2, \;1, 3/2\;
and \;2.

\bigskip

\textbf{(6)} A symmetry of the superspace is formed by a realization of the
generators with bosonic variables of the harmonic oscillator as Lagrangian.
Taking a line element corresponding to such superspace a physical state of
spin 2 can be obtained and related to the metric tensor.

\bigskip

\textbf{(7)} The metric tensor is discretized simply by taking the discrete
series given by the basic states (coherent states) in the number $n$
representation, consequently the metric tends to the classical (continuous)
value when $n$ $\rightarrow\infty$.

\bigskip

\textbf{(8)} The results of this paper have implications for the lowest
level of the discrete spectrum of space-time, the ground state associated to 
$n = 0$ and its characteristic length, in the black hole history of black
hole evaporation.

\bigskip

\textbf{(9)} Moreover, recently we have successfully applied this general approach in
physical scenarios of current interest obtaining coherent states of quantum
space-times for black holes and de Sitter space-time, in our Ref \cite%
{cirilo-sanchez}.

\newpage

\textbf{ACKNOWLEDGEMENTS}

\bigskip

\bigskip

DJCL acknowledges the institutions CONICET and the Keldysh Institute
of Applied Mathematics; and the support of the Moscow Center of
Fundamental and Applied Mathematics, Agreement with the Ministry of Science
and Higher Education of the Russian Federation, No. 075-15-2022-283.

\end{document}